\documentclass[12pt]{spieman}  
\usepackage{amsmath,amsfonts,amssymb}
\usepackage{graphicx}
\usepackage{setspace}
\usepackage{tocloft}
\usepackage{lineno}

\title{A New Method for Aperture Masking Interferometric Imaging: Demonstration with the JWST.}

\author[a]{Christopher L. Carilli}
\author[b]{Bojan Nikolic}
\author[c]{Laura Torino}
\author[d]{Nithyanandan Thyagarajan}
\author[c]{Ubaldo Iriso}

\affil[a]{National Radio Astronomy Observatory, Socorro, NM, USA, 87801}
\affil[b]{Astrophysics Group, Cavendish Laboratory, Univ. of Cambridge, Cambridge CB3 0HE, UK}
\affil[c]{ALBA - CELLS Synchrotron Radiation Facility\\Carrer de la Llum 2-26, 08290 Cerdanyola del Vallès (Barcelona), Spain}
\affil[d]{Commonwealth Scientific and Industrial Research Organisation (CSIRO), Space \& Astronomy, P. O. Box 1130, Bentley, WA 6102, Australia}

\cftpagenumbersoff{figure}
\cftpagenumbersoff{table} 
\begin{document} 
\maketitle

\begin{abstract}
We present a new method for aperture masking interferometric (AMI) imaging at near-IR wavelengths using radio astronomical techniques. The method starts with derivation of interferometric visibilities from a Fourier transform of the interferograms.  An iterative joint optimization process is then employed, using self-calibration of the interferometric element-based complex voltage gains (i.e. electric fields), and CLEAN deconvolution to obtain the source structure. We demonstrate the efficacy of the method using the NIRISS aperture masking interferometer on the James Webb Space Telescope (JWST) at 4.8~$\mu$m and 3.8~$\mu$m.  Due to a number of effects (the large pixel size, charge migration, near-field optics), the method also requires an initial visibility-based amplitude normalization using observations of a well know point-source calibration star. We employ early science observations of the dusty binary Wolf-Rayet star WR137. Images with a dynamic range (peak/rms) of $\sim 240$ on the target, and $\sim 1000$ on the calibrator, are synthesized from a short integration. The self-calibration process determines the photon path-lengths through the optical system to each aperture using data on the target source itself, thereby representing an essentially 'real-time', precise wavefront error sensor. Four independent measures of the JWST mirror segment pistons (two wavelengths for two sources), agree to within 10~nm to 15~nm, comparable to the expected errors based on an analysis of closure phases on the calibrator star.  Including a baseline-based phase correction improves the dynamic range of the final images by about 23\%.
\end{abstract}


{\noindent \footnotesize\textbf{*}Chris Carilli,  \linkable{ccarilli@nrao.edu} }


\section{Introduction}
\label{sect:intro}  

We report on application of new techniques for analysis of aperture masking interferometry at near-IR wavelengths using the NIRISS system on the JWST\cite{Sivaramakrishnan2023,lau2024,soulain2020,blakely2025}. These techniques are enabled by the excellent stability and sensitivity of the JWST. 

Our approach incorporates the techniques employed in radio astronomical interferometric imaging\cite{TMS2017,SIRA-II}. The process starts with derivation of interferometric visibilities (baseline fringe amplitudes and phases) from the interferograms, then an iterative process of model fitting (or deconvolution), and self-calibration of interferometric element-based complex gains to correct for hole-based phase and amplitude corruptions. The process generates an image of the source structure, a measure of the illumination pattern across the mask (gain amplitudes), and photon path-lengths through the optical system (gain phases). 

We have recently demonstrated the efficacy of these techniques in optical aperture masking interferometry in the laboratory at the ALBA synchrotron light source \cite{Nikolic2024,iriso2024,torino2025,thyagarajan2025}.  At ALBA, we recover the synchrotron source structure to 1\% accuracy, and the aperture illumination pattern to similar accuracy. The ALBA data were also employed as a precise wavefront sensor, deriving wavefront path-lengths to each interferometric element (hole in the mask), to sub-nm precision on millisecond timescales, thereby correcting for both static and time-variable phase fluctuations due to vibration of the optics and laboratory turbulence\cite{carilli2025}.

For the JWST, there is an added complication in that there are known baseline-based visibility amplitude corruptions due to the large pixel size (under-sampling of the fringe spacings), and the 'brighter-fatter' effect due to charge migration for bright sources\cite{argyriou2023}. Baseline-based corruptions are not corrected through self-calibration. In Section~\ref{sec:baseline}, we show that these effects are substantial, with up to a factor two deviation of closure amplitudes from expected values on the calibration star. Hence, a visibility-based coherence correction must first be made using measurements on the calibration star. Further, the pupil optics of AMI on the JWST may lead to potential Fresnel diffraction (near-field) baseline-based phase corruptions to the wavefront\cite{desdoigts2025}. We estimate the magnitude of these corruptions through analysis of closure phases on the calibrator star, and we make a final correction to the processing using the calibrator data as a measure of these errors (Section~\ref{sec:visphs}). 

We employ early science, Director Discretionary data on the dusty binary star WR137 \cite{lau2024}. This star system consists of newly formed dust from the colliding winds of the massive binary Wolf–Rayet star. These data have been previously analyzed thoroughly using the standard AMI closure phase and visibility amplitude techniques by Lau et al.\cite{lau2024}, as well as using a forward modeling approach to imaging and calibration\cite{charles2025}. These analyses provide an excellent point of comparison. We do not re-consider the physical processes of the binary star system, as discussed at length in Lau et al.\cite{lau2024}. In short, Lau et al. state that the JWST AMI  results are best explained by a: 'geometric colliding-wind model with dust production constrained to the orbital plane of the binary system and enhanced as the system approaches periapsis.' Note that the binary star separation is not resolved with the JWST AMI observations. 

The focus of this report is to demonstrate reliable reproduction using our new techniques of the images produced from previous standard processing methods, and explore other aspects of the JWST system that can be probed through self-calibration techniques.  For completeness, we start with an overview of basic concepts in interferometry. We then discuss the data processing to visibilities, and the derivation of closure quantities. The self-calibration and imaging results for the source are then presented, as well as measurements of the JWST mirror segment pistons. Lastly, we consider visibility-based phase corrections. 

We emphasize that there are a number of standard JWST-AMI reduction pipelines, including standard closure phase + visibility amplitude model fitting\cite{lau2024,Sivaramakrishnan2023,blakely2025}, and a recent full Bayesian forward-modeling approach to image reconstruction using a detailed telescope model\cite{desdoigts2024,desdoigts2025}, the latter of which achieved a source-planet dynamic range of 9mag (brightness ratio = 4000). Early (pre-launch) modeling of JWST -AMI data also considered Fourier imaging with visibilities and CLEAN deconvolution, although the nature of the interferometric errors (i.e. the relative contribution of closing and non-closing errors), were not well quantified at the time, so self-calibration could not be incorporated reliably into the modeling\cite{ford2014}. 

In general, image uniqueness is a well known uncertainty when imaging with interferometers with sparse Fourier plane coverage\cite{TMS2017}. In such cases, multiple data reduction and image restoration algorithms are critical for improving confidence in the final result. A recent case-in-point being the imaging of the general relativistic shadow of the super-massive black hole in M87 with the Event Horizon Telescope at 230 GHz, for which multiple forward-modeling approaches, as well as standard Fourier imaging and deconvolution methods, have been employed in order to lend confidence to the observed 'ring' on scales of a few gravitational radii\cite{EHT,carilli2023}.

\section{Interferometric (Fourier) Imaging, Self-Calibration, and Closure}
\label{sec:method}

Interferometric imaging and self-calibration techniques have been well documented in the context of radio astronomical measurements\cite{TMS2017}. We summarize the basics for completeness (material can also be found in Carilli et al.\cite{carilli2025}).

\subsection{Self-calibration}
\label{sec:selfcal}

The spatial coherence (or \textit{visibility}), $V_{ab}(\nu)$,
corresponds to the cross correlation of two quasi-monochromatic voltages centered at frequency $\nu$ of the same polarization sensed by two spatially distinct elements in the aperture plane of an interferometer. The visibility relates to the intensity distribution of an incoherent source in the far field, $I(\hat{\mathbf{s}},\nu)$, via the van Cittert-Zernike theorem  \cite{vanCittert34, Zernike38}, which after certain approximations reduces to a Fourier transform \cite{TMS2017,Born+Wolf1999}:

\begin{align}
    V_{ab}(\nu) &= \int_\textrm{source} A_{ab}(\hat{\mathbf{s}},\nu) I(\hat{\mathbf{s}},\nu) e^{-i2\pi \mathbf{u}_{ab}\cdot \hat{\mathbf{s}}} \mathrm{d}\Omega \ = \gamma_{ab}(\nu)e^{i\phi_{ab}(\nu)}, , \label{eqn:VCZ-theorem}
\end{align}

\noindent where, $a$ and $b$ denote a pair of array elements (eg. holes in a mask), $\hat{\boldsymbol{s}}$ denotes a unit vector in the direction of any location in the image, $A_{ab}(\hat{\mathbf{s}},\nu)$ is the spatial response (the `power pattern') of each element (in the case of circular holes in the mask, the power pattern is the Airy disk, with a size inversely proportional the hole diameter), $\mathbf{u}_{ab}=\mathbf{x}_{ab} (\nu/c)$ is the ``baseline'' vector = the vector spacing ($\mathbf{x}_{ab}$) between the element pair in units of wavelength (or inverse radians), and $\mathrm{d}\Omega$ is the differential solid angle element on the image (focal) plane. The coherence, or amplitude, of the visibility is $\gamma_{ab}$, and the phase is $\phi_{ab}$.

In practice, a visibility is a complex-valued measurement in the aperture plane defined by $u$--$v$ coordinates, which, after Fourier transform, corresponds to a sinusoidal (`fringe') pattern in the image-plane, with a spatial frequency and orientation determined by the baseline vector, and having an amplitude corresponding to the power of the source mutual coherence at that fringe spacing ($\gamma_{ab}$), and a phase corresponding to the position of that fringe relative to the adopted phase center ($\phi_{ab}$). This concept of a sinusoidal fringe in an image from an interferometric baseline has been the basis of interferometry since Young's original 2-slit experiment of 1803\cite{young1803}.

The JWST employs optical aperture masking interferometry, in which a mask with a specific pattern of holes, or interferometric elements, is placed in the pupil wheel. The JWST mask system includes a 7-hole non-redundant aperture mask\cite{Sivaramakrishnan2023}. Non-redundancy means that each baseline in the mask (vector hole separations), is unique, thereby avoiding decoherence of redundant visibilities due to phase fluctuations across the full aperture. 
An image of the interference pattern of the mask, or interferogram, is generated on the NIRISS camera. Interferometric visibilities are generated by a Fourier transform of the interferograms (Section~\ref{sec:measurements}).

The measured visibilities can be corrupted by distortions introduced by the propagation medium or the relative illumination of the holes, or other effects in the optics (vibrating or non-perfect mirrors or lenses). For the JWST, the atmosphere is clearly not an issue, but optical effects remain. In many cases, these distortions can be factorized into multiplicative element-based complex voltage gain factors, $G_a(\nu)$ and $G_b(\nu)$, where $a$ and $b$ represent the two interferometer elements in the visibility baseline ($G = A_Ge^{i\phi_G}$, where $A_G$ and $\phi_G$ are the amplitude and phase of the complex gains per element). Thus, the corrupted measurements, $V_{ab}^\prime(\nu)$, are given by the complex product:
\begin{align}
    V_{ab}^\prime(\nu) &= G_a(\nu) \, V_{ab}(\nu) \, G_b^\star(\nu) \, , \label{eqn:uncal-vis}
\end{align}
\noindent where, $\star$ denotes a complex conjugation, and $V_{ab}(\nu)$ is the true source visibility. 

The process of interferometric self-calibration determines these complex voltage gain factors in parallel with determining source structure. A physically reasonable starting model for the source is assumed, $V_{ab}(\nu)$. Using the measurements, $V_{ab}^\prime(\nu)$, Equation~(\ref{eqn:uncal-vis}) is then inverted to derive the complex voltage gains, $G_a(\nu)$, using an optimization criterion, such as least squares fitting \cite{Schwab1980, Schwab1981, Readhead+Wilkinson1978, Cornwell+Wilkinson1981, Nikolic2024}. A new source model is then derived from the gain-corrected visibilities through model fitting or imaging (Fourier inversion) and deconvolution, and the process is iterated until convergence. A block diagram of the self-calibration process can be found in Carilli et al.\cite{carilli2025}.

The self-calibration process converges since there are typically more measurements ($[N(N-1)]/2$ complex cross-correlations, or visibilities, yielding $N(N-1)$ real-valued measurements, where $N$ is the number of interferometric elements), than fitted parameters ($2N-1$ real parameters of $N$ element-based complex gains with one parameter for the inconsequential absolute phase removed, and model parameters), at least for reasonably sampled visibility data and a not-too-complex source \cite{Schwab1980, Schwab1981}. 

The complex voltage gains encode critical information about the optical system.  The amplitude gains encode the aperture plane illumination pattern. The hole-based phase gains, are linearly proportional to wavefront path-length delays to the aperture, $\rm \delta L(t)$, relative to the reference hole\footnote{Phase, or path-length delay, is always a difference measurement with respect to a reference position since the wavefront is comprised of many incoherent photons, for which absolute phase is ill-defined, but phase difference between spatial positions in the mask remains a well defined and physical quantity common among photons, thereby allowing for mutual coherence and application of the van Cittert-Zernike theorem.}, through the simple relation: 

\begin{align}
\rm \delta L = \lambda \times (\phi_G /360^\circ), \label{eqn:path}
\end{align}

\noindent where $\rm \phi_G$ is in degrees. Hence, the gain phases correspond to an accurate (small fraction of a wavelength), high time resolution wavefront sensor\cite{carilli2025}. 

\subsection{Closure Quantities}
\label{sec:intbaseline}

An alternative to self-calibration to correct for element-based corruptions in interferomerty is to use 'Closure quantities'. Closure quantities are products of visibilities that are robust to element-based voltage phase and amplitude corruptions\cite{Jennison1958,Pearson+Readhead1984,Pearson+1981, Wilkinson+1977,Thyagarajan+Carilli2022,Thyagarajan+2022, Samuel+2022,thyagarajan2025}.

Closure phase is defined as the argument on a complex product of visibilities from a closed triad of three aperture elements, for which the argument becomes the sum of the measured phases
(primed quantities), of the three visibilities\cite{Jennison1958,TMS2017,Readhead+Wilkinson1978,Cornwell+Fomalont1999,Thyagarajan+Carilli2022}:

\begin{align}
    \phi_{pqr}' &= \phi_{pq}' + \phi_{qr}' + \phi_{rp}' = \phi_{pqr}^\textrm{source} \label{eqn:Clph}
\end{align}
\noindent  From Equation~(\ref{eqn:uncal-vis}), $\phi_{pq}' = \phi_{pq}^\textrm{source} + \phi_G^p - \phi_G^q$, etc... It can be seen that the element-based phase corruptions subtract, such that the measured closure phase equals the true source closure phase, independent of element-based phase corruptions. The manifestation of closure phase in the image plane as a consequence of 'shape-orientation-size' conservation, can be found in Thyagarajan \& Carilli\cite{Thyagarajan+Carilli2022}.

The closure amplitude is defined as a visibility product of amplitudes on a closed loop of four aperture elements, hereafter referred to as a \textit{quad}, as \cite{TMS2017,SIRA-II,thyagarajan2025}:
\begin{align}
    \gamma_{pqrs}' &= \frac{\gamma_{pq}'\gamma_{rs}'}{\gamma_{ps}'\gamma_{rq}'} = \gamma_{pqrs}^\textrm{source}, \label{eqn:ClAmp}
\end{align}
\noindent Again, from Equation~(\ref{eqn:uncal-vis}), $\gamma_{pq}' = A_pA_q\gamma_{pq}^\textrm{source}$, and  it can be seen that element-based gain amplitude corruptions normalize-out, and the closure amplitude, $\gamma_{pqrs}$, is independent of the hole illuminations. 

Most interferometric imaging in optical astronomy employs a process of model fitting to the closure phases and visibility amplitudes, including analyses of AMI data from the JWST\cite{lau2024, blakely2025, Haniff2007,Baldwin+1986}. Occasionally, higher order visibility combinations are employed, such as Kernel phases \cite{Martinache+2020}. Self-calibration has not been used in ground-based optical astronomical interferometry due to the paucity of photons per coherence time. Closure phase allows for integration over many coherence times, and hence build up of the S/N required for model fitting \cite{Buscher2015,Haniff2007,Baldwin+1986}.  Moreover, self-calibration entails correcting amplitudes and phases of the voltage response (i.e. electric fields), at each element. This is a natural working space for radio interferometers, since voltages are coherently amplified at each element prior to cross correlation to visibility power. Optical interferometry fundamentally works in the image plane by making images of interferometric fringes, i.e. power = voltage$^2$, and the advantages of going back to the voltage plane have not been as obvious.

By making only element-based corrections, the self-calibration process preserves closure quantities. When the signal is adequate, there are a few advantages to working directly with visibility phases and amplitudes as opposed to closure phases and visibility amplitudes. The most obvious advantage is the greater number of independent constraints: the number of visibility phases is $N(N-1)/2$, while the  number of close phases is $[N(N-1)/2 - (N-1)]$. So for a 7-element interferometer, there are 21 independent visibility phase measurements vs. 15 closure phases.  Moreover, closure phase is essentially just a measure of the source symmetry properties\cite{TMS2017, Thyagarajan+Carilli2022}. For instance, any point-symmetric source brightness distribution will have zero closure phase. This is not necessarily true for visibility phases, which retain information on the relative positions of source structure. 

The self-calibration techniques outlined herein are also essentially 'real-time' corrections of wavefront errors.\footnote{The use of a non-redundant mask allows for this post-facto but essentially real-time correction of wavefront errors, since coherence is maintained for each independent visibility fringe.} While the JWST is very stable, there are small time dependent changes in the full mirror, and mirror segments, due to thermal gradients and structural stress, for which time dependent wavefront correction on the source itself can improve imaging performance\cite{defurio2025}.

Lastly, working directly with visibilities instead of closure quantities allows employment of years of development of Fourier imaging and point-spread-function (PSF) deconvolution algorithms, such as CLEAN \cite{clark1980, hogbom1974}. The CLEAN algorithm in particular is relatively source structure agnostic, using point source basis functions. CLEAN has been demonstrated to recover very complex source structure with adequate Fourier plane, or $u,v$ coverage \cite{Perley1999}. 

\subsection{Non-closing (baseline-based) corruptions}
\label{sec:nonclosing}

Fundamental to self-calibration and closure analysis is the assumption that the optical corruptions can be factorized into interferometric element-based terms as described in Equation~(\ref{eqn:uncal-vis}). If there are corruptions that are idiosyncratic to a interferometric baseline, then both methods will be limited\cite{Perley1999}. Analysis of closure quantities on the calibrator give a quantitative measure of the magnitude of the non-closing errors (Section~\ref{sec:baseline}). 

Baseline-based errors can be addressed if one observes a strong, point source calibrator, in which case the target source visibilities can be corrected by measurements on the calibrator.  A potential danger with baseline-based visibility correction is that the process is under constrained, and hence, it becomes trivial to simply `turn the source into the model' \cite{Perley1999}. This is clearly the case for self-calibration, and baseline-based self-calibration should be avoided. There is less risk when using a known, bright point source calibrator to derive baseline-based corrections, and then applying these to the target. However, one then must assume there is no change in the system response between source and calibrator observations. Overall, element-based corrections are preferred, except in the case when there are known non-closing errors and a bright, point source calibrator is available. 

For JWST, there are a few known sources of non-closing errors. Considering the visibility amplitudes, the NIRISS pixel size is a substantial fraction of the longest fringe spacing, hence under-sampling the PSF. Further, charge migration on bright sources leads to fringe broadening that will be baseline and source flux dependent (`brighter-fatter' effect  \cite{argyriou2023,desdoigts2024}). 

There are also phase corruptions due to the JWST AMI pupil-plane optics, which may also couple into the amplitude response. The system is close to the Fresnel diffraction limit (near-field), which could affect the measured visibilities, since interferometric imaging (the van Cittert-Zernike theorem) assumes far-field optics.\cite{desdoigts2025} 

Desdoigts et al.\cite{desdoigts2025} have performed an in-depth analysis of these phenomena at the JWST using a forward-modeling approach involving a detailed model for the JWST optics, and using a point-source calibrator to quantify the effects. In our analysis below, we consider these non-closing effects on the visibilities using the calibrator star as a reference (Section~\ref{sec:measurements}, Section~\ref{sec:baseline}, Section~\ref{sec:visphs}).  We quantify the level of corruption using the closure quantities, and demonstrate application of baseline-based corrections for both amplitude and phase. 

\section{Observations and initial data processing to visibilities}

\subsection{Measurements}
\label{sec:measurements}

The observations were taken during JWST early science in program DD-ERS program 1349 \cite{lau2024}. In brief, the aperture masking interferometry mode of the NIRISS instrument was employed \cite{Sivaramakrishnan2023,soulain2020}. Observations of WR 137 and a PSF calibrator star (HD 228337) were made on July 15, 2022, using the F380M (wavelength = 3.825~$\mu$m, width 0.205~$\mu$m), and F480M (4.815~$\mu$m, 0.298~$\mu$m) filters. The standard AMI observing parameters were employed, using the “NISRAPID” readout pattern (1 integration per group), with the SUB80 (80 ×80 pixel) subarray. The NIRISS detector plate scale is 65 mas/pixel. The readout time per frame is 75.44 ms, which equals the group time in this readout mode. At 4.8~$\mu$m, four groups were averaged to get 0.30~s independent integrations in the final interferogram timeseries cube. At 3.8~$\mu$m, 2 readouts were averaged to get 0.15~s independent integrations in the final interferogram timeseries cube. The total and effective exposure times of the F480M and F380M observations of WR 137 were 10.6 and 8.0 min, respectively,  and 11.2 and 6.8 min at F380M. The total exposure time on HD 228337 was the same to that of WR 137. 

The total flux density of WR137 at 4.8~$\mu$m is 2.8~Jy \cite{lau2024}, however, in the analysis below we deal with relative intensities since we are only interested in testing the imaging and self-calibration methodology and dynamic range. We do not re-address the physical processes involved in the dusty binary star system.  We analyze image data from the MAST archive data corresponding to the time series image cube with calibration and cosmic ray excision applied (`calints' FITS timeseries cubes). 

\begin{figure}[!htb]
\centering 
\centerline{\includegraphics[scale=0.29]{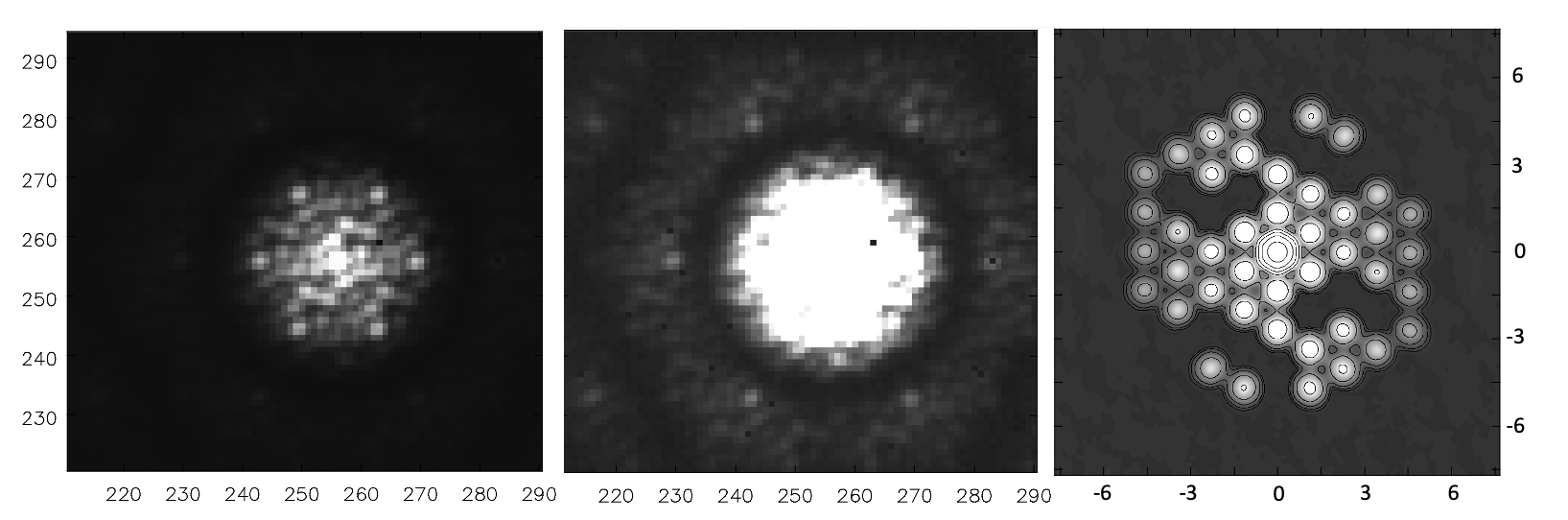}}
\caption{Left and Center: example JWST AMI interferogram of WR137 at 4.8$\mu$m. The pixel size is 65mas. The grayscale stretch of the center image is saturated to emphasize the first diffraction lobe of the power pattern (`primary beam') set by the size of the hexagonal apertures of the JWST aperture mask. Right: The amplitudes of the u,v samples derived by a Fourier transform of the AMI image. X and Y axes are the interferometer u,v baseline coordinates, in meters. 
}
\label{fig:Im.UV}
\end{figure}

Figure~\ref{fig:Im.UV} shows an example interferogram at 4.8~$\mu$m, plus the visibilities from the Fourier transform of this interferogram. In radio astronomical interferometry, this image would be called the `dirty image', meaning the source brightness convolved with the point spread function of the 7-hole mask. Clearly evident on the image are the PSF grating lobe peaks due to the regular spacings of the apertures of AMI. Also seen is the power envelope of each aperture (in radio astronomy: the `primary beam'). For the JWST, the apertures are hexagons (see Figure \ref{fig:piston}), and hence the power pattern is roughly, but not exactly, an Airy disk.

\subsection{Visibilities and sampling}
\label{sec:vis}

Interferometric complex visibilities were generated via a Fourier transform of the interferograms (Figure~\ref{fig:Im.UV}).  Prior to transforming, the interferograms are centered on the peak brightness position derived by image-plane fitting. The 7-hole non-redundant mask (Figure~\ref{fig:piston}) produces 21 visibilities (plus their Hermitian conjugates). Again, each visibility corresponds to one interferometric fringe in the image plane, with a sinusoidal fringe spacing and orientation set by the vector baseline, a strength given by the visibility amplitude (one Fourier component of the source brightness), and a position relative to a reference position dictated by the visibility phase. 

The size of each u,v-sample is set by the Fourier transform of the primary beam, as dictated by the hole shape. The visibility amplitudes and phases were derived from the resulting Fourier component images through a complex weighted sum over the Real and Imaginary visibility sample area, integrating down to the $\sim 30$\% point relative to the peak of the sample. The details of the process can be found in Nikolic et al.\cite{Nikolic2024}.

\begin{figure}[!htb]
\centering 
\centerline{\includegraphics[scale=0.28]{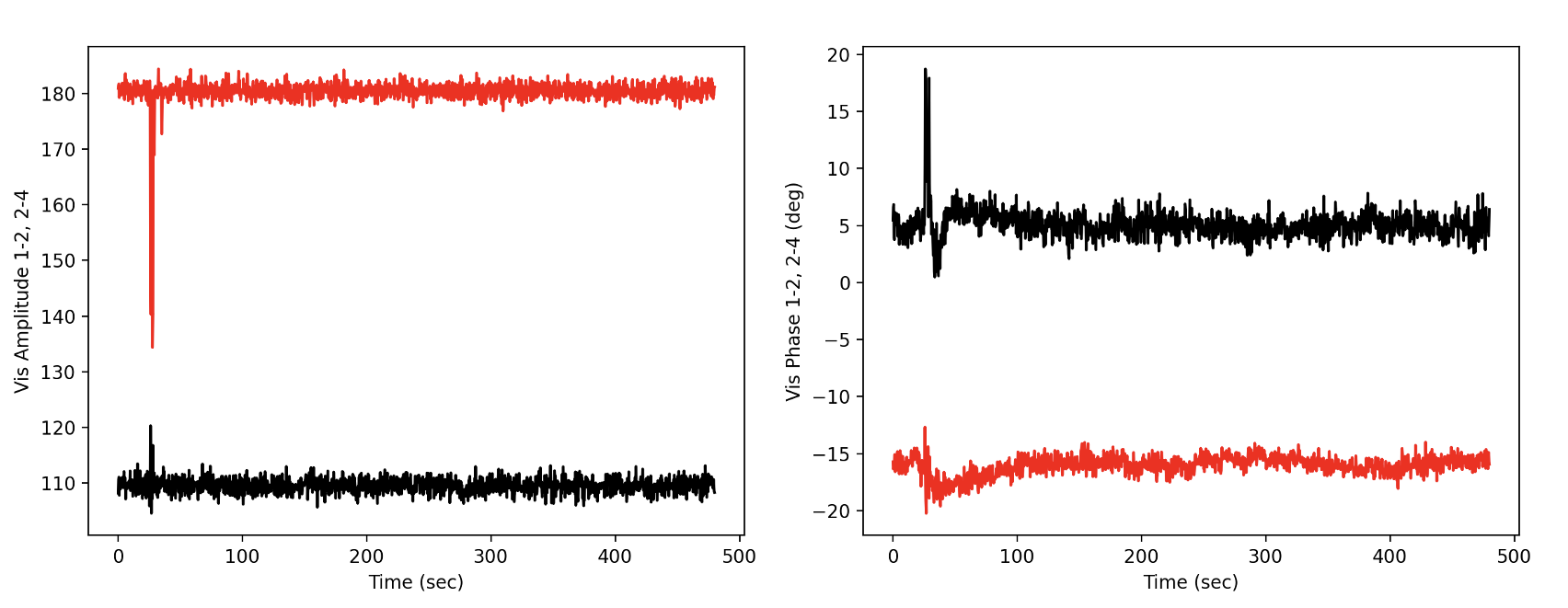}}
\caption{Example time series of visibility amplitude  and phase for two baselines (1-2 red; 2-4 black) on WR137 at 4.8$\mu$m. The integration time per point is 0.3~s. 
}
\label{fig:VisAP}
\end{figure}

Figure~\ref{fig:VisAP} shows examples of the time series of visibilities at 0.30~ms integration time for the 4.8~$\mu$m data on WR137 for two baselines. The amplitudes and phases are very stable (rms amplitudes $\sim 1\%$, rms phases $\sim 1^\circ$), except for the first 200 or so integrations at 4.8~$\mu$m on WR137.  Inspection of the image cube for this time range shows clear artifacts immediately around the bright peak, resembling cosmic ray blanking. This time range was left out of the analysis. For the other sources, there were a few bad individual integrations which are easily excised (or even ignored). Given this stability, in our analysis below, we boxcar average the time series to 12~s.

\section{Measured baseline-based corruptions}
\label{sec:baseline}

There are a number phenomena with AMI on the JWST that could lead to baseline-based, and therefor, non-closing errors, thereby invalidating equation~\ref{eqn:uncal-vis} and hence assumptions in  closure quantity analysis, or element-based self-calibration\cite{argyriou2023,desdoigts2024,desdoigts2025}. 

For amplitudes, one known problem with AMI on JWST is under-sampling of the spatial fringes due to the NIRISS pixel size of 65~mas. For AMI, the longest baseline is 5.3m, implying a fringe spacing (peak to peak) of $\lambda/B = 200$~mas at 4.8~$\mu$ and 150~mas at 3.8~$\mu$m. So at most, the fringe is sampled by two or three pixels per full fringe wavelength. There is a well known relationship between decoherence and sampling: Coherence $\rm = sinc(X)$, where $\rm X = \pi \times (pixel~size/fringe~spacing)$. 

A second phenomena that may cause decoherence relates to source brightness and charge migration or peak charge loss (`brighter-fatter' effect\cite{argyriou2023,desdoigts2024}), although, although for this dataset, Lau et al.\cite{lau2024} state: 'these data... were found to display non-linearity due to charge migration of at most 2\% at the highest signal level reached. The impact of charge migration on our dataset is therefore negligible.'  It is also possible that the pupil plane near-field optics of AMI at the JWST (see below), contributes to amplitude corruptions\cite{desdoigts2025}.

\begin{figure}[!htb]
\centering 
\centerline{\includegraphics[scale=0.2]{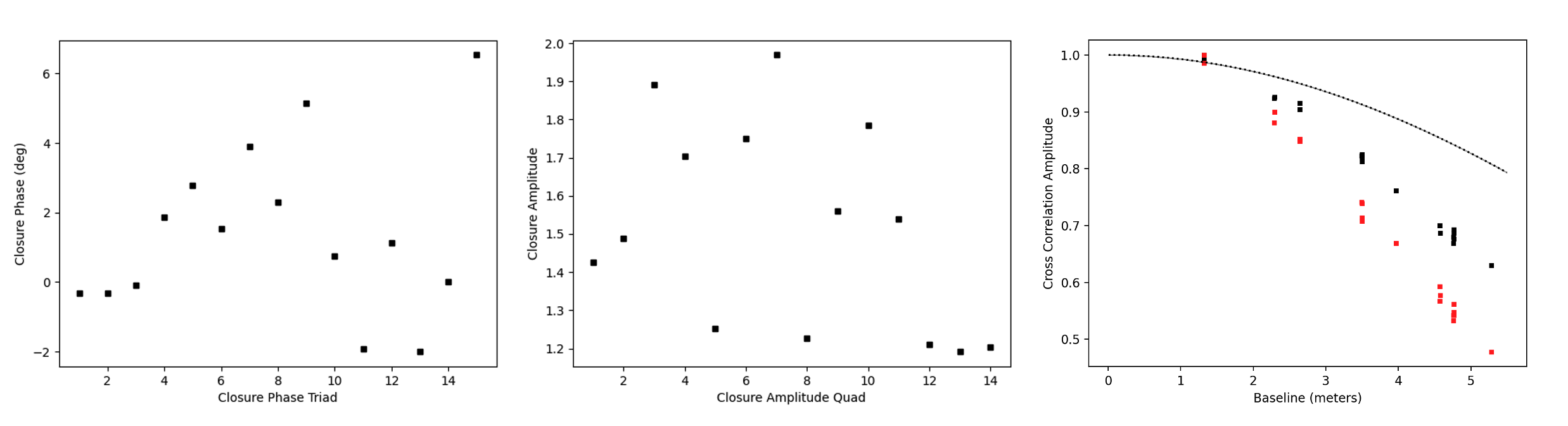}}
\caption{{\bf Left:} Closure phases for all the independent closure triads in the mask, on the calibration star HD 228337. For a point source, the closure phase should be zero, unless there are baseline-dependent phase errors.  The X-axis numbers map to triads as: 1 = (0, 1, 2), 2 = (0, 1, 3), 3 = (0, 1, 4), 4 = (0, 1, 5), 5 = (0, 1, 6), 6 = (0, 2, 3), 7 = (0, 2, 4), 8 = (0, 2, 5), 9 = (0, 2, 6), 10 = (0, 3, 4), 11 = (0, 3, 5), 12 = (0, 3, 6), 13 = (0, 4, 5), 14 = (0, 4, 6), 15 = (0, 5, 6).
{\bf Center:} closure amplitudes for all the independent closure amplitudes quads in the mask on HD 228337. For a point source, the closure amplitudes should be unity. Departure from unity implies non-closing errors. The X-axis numbers map to quads as: 1 = (0, 1, 2, 3), 2 = (0, 1, 3, 4), 3 = (0, 1, 4, 5), 4 = (0, 1, 5, 6), 5 = (1, 2, 3, 4), 6 = (1, 2, 4, 5), 7 = (1, 2, 5, 6), 8 = (1, 2, 6, 0), 9 = (2, 3, 4, 5), 10 = (2, 3, 5, 6), 11 = (2, 3, 6, 0), 12 = (3, 4, 5, 6), 13 = (3, 4, 6, 0), 14 = (4, 5, 6, 0).
{\bf Right:} The normalized measured visibility amplitude on HD 228337 at 4.8$\mu$m (black) and 3.8$\mu$m (red) vs. baseline length. The dotted line shows the model for expected decoherence due to the large pixel size and under-sampling of the spatial fringes at 4.8$\mu$m (Section~\ref{sec:measurements}). 
}
\label{fig:Closure}
\end{figure}

Figure~\ref{fig:Closure} shows the normalized visibility amplitudes on the calibration star HD 228337 at 4.8~$\mu$m and 3.8~$\mu$m. There is a clear decrease of visibility amplitude with baseline length to 50\% on longest baselines at 3.8~$\mu$m. The star itself is effectively a point source (smaller than 0.5~mas in HD catalog), so there should be full (unit) coherence for all baselines. Figure~\ref{fig:Closure} also shows the expected decoherence with baseline length based on the pixel scale and using the relationship between pixel size and fringe spacing above. The measured coherences are much lower than expected from just under-sampling. 

This baseline-dependent decoherence is further demonstrated in the closure amplitudes on HD 228337, as shown in Figure~\ref{fig:Closure}. Closure amplitudes on an unresolved source should be unity.  But in this case, all values are $> 1.2$, and range up to almost two. 

Regardless of origin, these large corruptions to coherence with baseline length require baseline-based correction.  In the analysis below, we make an empirical correction to visibility amplitudes using the measured decoherence on the calibrator star. Note that the Brighter-Fatter effect will be flux dependent, so it is desirable to have similar magnitudes for the target source and calibrator. Fortunately, this is the case for WR137 and HD 228337, which are both of order M magnitude $\sim 5$.

We note that a Gaussian fit to the uncorrected visibilities for HD 228337 yields an apparent source size of 70~mas, which is very close to the pixel size. 

Considering phases, Desdoigt et al.\cite{desdoigts2025} present the possibility of non-closing phase errors due to Fresnel diffraction (near-field) optics for AMI in the pupil plane. Figure~\ref{fig:Closure} shows the closure phases for all independent triads in the mask on HD 228337. For an unresolved source, closure phases should be zero. The measured closure phases on HD 228337 range up to $6.5^\circ$, with a mean absolute value of $\sim 2^\circ$, and an rms $\sim 2^\circ$. While this is a relatively small corruption (1\% to 2\% of a full turn of phase), such small phase errors could be a limiting factor when performing very high dynamic range measurements, such as searching for planets. For example, Perley\cite{Perley1999} calculates the expected image dynamic range for an N element array given random element-based phase errors: $\rm DNR= N/(\sqrt{2} \times \delta\phi_G)$, where $\delta\phi_G$ are the element-based phase errors in radians (Perley equ. 13-8). Hence, to reach an image dynamic range of 1000 with a 7 element array would require residual element-based phase errors $\le 0.3^\circ$.

In the analysis below, we first proceed with self-calibration without correcting for these small non-closing phase errors. Such an analysis then still allows for consideration of wavefront sensing and mirror segment piston phase measurements (Section~\ref{sec:wavefront})\footnote{Baseline-based phase corrections using the calibrator data remove both the closing (piston phase) terms and the non-closing common-mode optics terms.}. We then correct the baseline-based errors by subtracting the measured phases on the calibration star from those of the source, and measure the improvement in the final images (Section~\ref{sec:visphs}). 

\section{Results and Analysis} 

\subsection{Imaging and self-calibration}
\label{sec:imaging}

For the self-calibration process, we employ visibilities on WR137 which have had the amplitudes normalized by the coherences measured on the calibration star, as shown in Figure~\ref{fig:Closure}.  The process starts with a point source model, with phase-only self-calibration for all datasets. A new source model is then generated  from the self-calibrated data using CLEAN deconvolution to obtain a CLEAN components model. The task TCLEAN in CASA was employed for Fourier inversion and deconvolution, using natural weighting of the visibilities, a pixel size of 5~mas, and an image size of 256x256 pixels. The fitted Gaussian FWHM of the interferometer PSF by TCLEAN (the Fourier transform of the Fourier plane coverage = $u,v$ sampling), was $0.13''$ at 4.8~$\mu$m and $0.10''$ at 3.8~$\mu$m. Two iterations of both amplitude and phase self-calibration were then performed, using the updated CLEAN component models. 

\begin{figure}[!htb]
\centering 
\centerline{\includegraphics[scale=0.28]{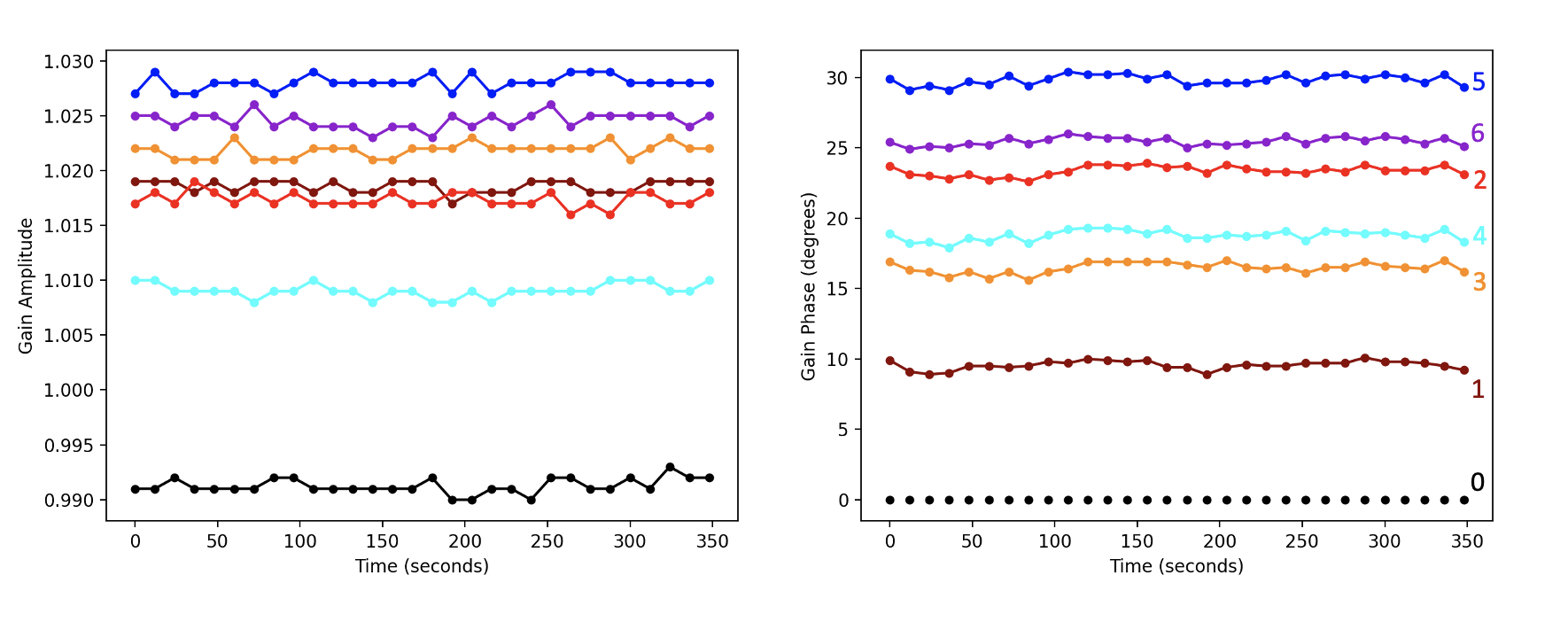}}
\caption{Time series of aperture-based amplitude  and phase self-calibration gain solutions on WR137 at 4.8~$\mu$m. The color-coding to aperture number is shown on the right plot (see Figure~\ref{fig:piston}). Note the zero start time is roughly time 100s in Figure~\ref{fig:VisAP}.
}
\label{fig:Gains}
\end{figure}

The time series of element-based phase and amplitude gain solutions are shown in Figure~\ref{fig:Gains} at 4.8~$\mu$m on WR137. The mean values in amplitude voltage gain range from 1\% to 3\%. These represent the residual differences between measurements on the star vs. the calibrator, given the baseline-based amplitude normalization discussed above. 

The mean phase gains range up to $30^\circ$. Below we show that these mean phases are likely due to small wavefront tilts originating in the source centering process prior to Fourier transforming (Section~\ref{sec:wavefront}).  The stability is excellent, with amplitude gain rms $\sim 0.1\%$ and phase gain rms $\sim 0.6^\circ$. There are slow systematic variations in the phase on timescales 30~s to 100~s at the level of $\sim 2^\circ$.

\begin{figure}[!htb]
\centering 
\centerline{\includegraphics[scale=0.35]{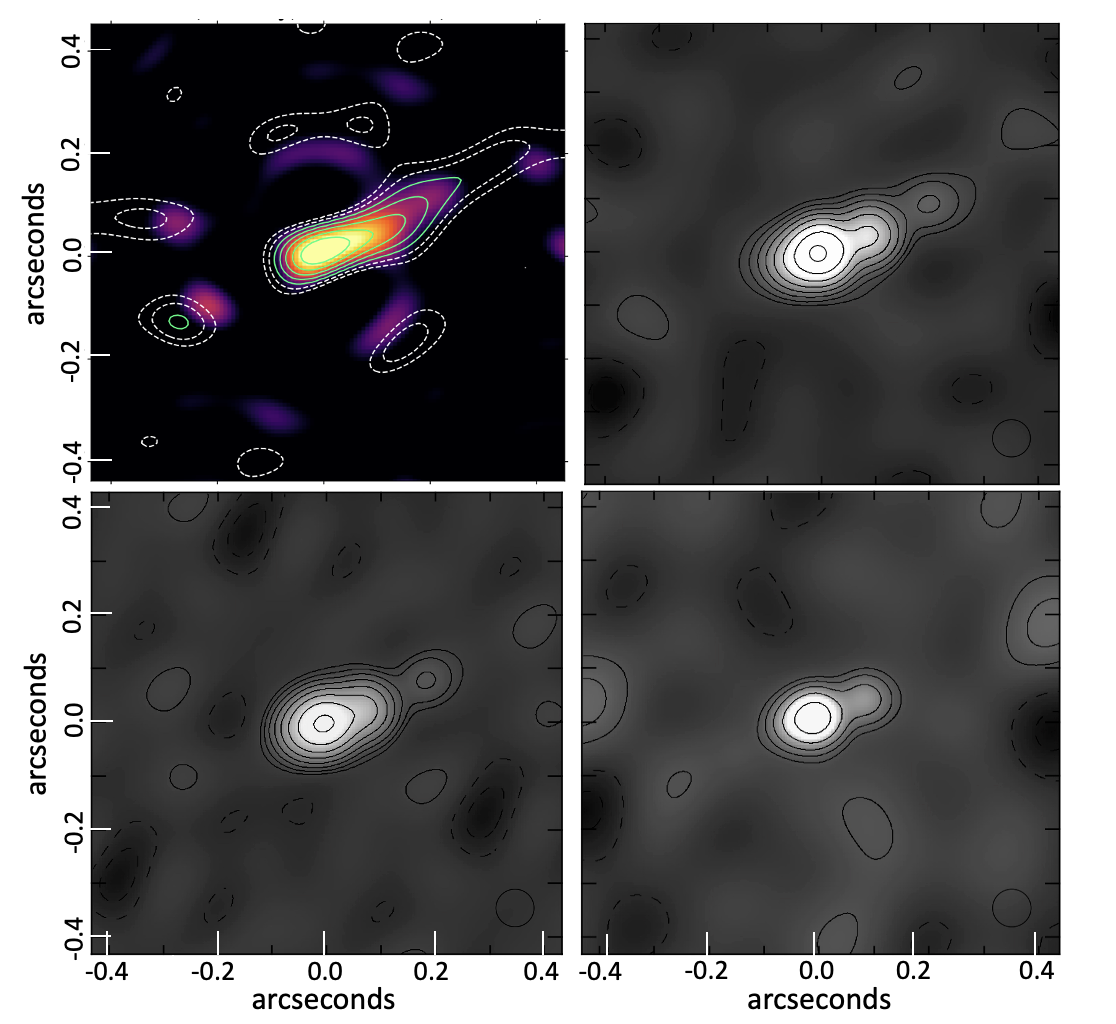}}
\caption{Top left: Image of WR137 from Lau et al.\cite{lau2024} (their figure 5). Contours are at 4.8$\mu$m and color scale is 3.8$\mu$m, with the first solid green contour = 0.33\% of the peak intensity. Top right: The self-calibrated and CLEANed image of WR137 at 4.8$\mu$m. Bottom left:  The self-calibrated and CLEANed image of WR137 at 3.8$\mu$m. In both cases, the contour levels are a geometric progression in factors of two, with the first level being 0.45\% of the peak surface brightness. Dashed contours are negative. In both cases the Gaussian restoring CLEAN beam has FWHM = 70~mas.  Bottom right: CLEAN WR137 image at 4.8$\mu$m without self-calibration. In this case, the first contour level is 2.7\% of the peak surface brightness.
}
\label{fig:4images}
\end{figure}

Figure~\ref{fig:4images} shows the final deconvolved images of WR137. Also  reproduced for reference are the images at 4.8~$\mu$m and 3.8~$\mu$m from Lau et. al.\cite{lau2024}. Note that a correction was made for telescope roll angle, based on information provided in Figure 1 in Lau et al.\cite{lau2024}. Images produced by summing all interferograms and processing, or processing each 12s integation separately and then summing the images, produced essentially identical results, thereby demonstrating the stability of phases and amplitudes, as shown in Figure~\ref{fig:VisAP} and Figure~\ref{fig:Gains}.

Given the short total integration and very limited $u,v$ coverage, the agreement between our images, and those produced by Lau et al.\cite{lau2024} and Charles et al.\cite{charles2025}, is encouraging, particularly given the very different analysis methods. The general structure and extent of the source is reproduced, including evidence for the fainter 300~mas filament noted by Lau et al. and Charles et al.\cite{lau2024, charles2025}, at a similar brightness level (relative to the peak), for the two very different data reduction procedures.

Values for the image parameters are listed in Table~\ref{table:image}. Again, we do not consider absolute flux densities, but only relative flux densities, given that the most relevant quantities for image quality analysis are the image dynamic = (peak surface brightness)/(off-source rms surface brightness), and the ratio of (peak surface brightness)/(peak 'image artifact'), where the peak artifact is measured as the peak off-source negative surface brightness. 

For WR137, the image dynamic range is 216 at 4.8~$\mu$m and 192 at 3.8~$\mu$m. For the point source calibrator HD 228337 the values are 1123, and 572, respectively. Our dynamic range on WR137 appears comparable to the best imaging results of Lau et al.\cite{lau2024}, presented in their Figures 5 and 7. 

Figure~\ref{fig:4images} also shows a CLEAN deconvolved image without self-calibration at 4.8~$\mu$m. In this case, the rms increases by a factor of six relative to the self-calibrated image, and the lower surface brightness 300~mas extension cannot be seen (Table~\ref{table:image}). 

\begin{table}
\centering
\footnotesize
\caption{Relative Brightness Values}
\begin{tabular}{lcccccc} 
  \hline
  \hline
Object & $\lambda$  & Total & Peak & rms & min & J2 \\
~ & $\mu$m  & counts & cnts/beam & cnts/beam & cnts/beam  & cnts/beam \\
\hline
WR137 & 4.8 & 201 & 134 & 0.62 & -2.1 &  4.1 \\
WR137 & 3.8 & 313 & 205 & 1.1 &  -3.7 &  6.3 \\
WR137 no self-cal & 4.8 & 197 & 137 & 3.6 & -12.4 & 0.23 \\
HD228 & 4.8 & 59.5 & 59.5 & 0.051 & -0.14  & -- \\
HD228 & 3.8 & 188 & 189 & 0.33 & -1.2 & --  \\ 
WR137 Complex$^a$ & 4.8 & 199 & 133 & 0.51 & -1.8 &  4.0 \\
WR137 Complex & 3.8 & 309 & 203 & 0.89 & -2.7 &  6.4 \\
\hline
\hline
\vspace{0.1cm}
\end{tabular}
\\ $^a$ Complex implies inclusion of baseline-based phase corrections using calibrator data. 
\label{table:image}
\end{table}

\subsection{Wavefront sensing: JWST segment piston metrology}
\label{sec:wavefront}

We next consider the metrology of the JWST mirror segments based on the phase self-calibration measurements. Again, the photon path-lengths through the optics are linearly related to the wavelength and the phase gain solutions (equ.\ref{eqn:path}). Our analysis of segment pistons first entailed converting gain phases to path-lengths using this linear relationship. We then fit a simple planar wavefront to remove the lowest order term. This tip-tilt term is likely dominated by image centering prior to Fourier transform: even a small shift in the image centering process will lead to a substantial apparent wavefront tilt equal to the offset in the image-plane from center\cite{carilli2025}. Centering accuracy is limited by the large pixel size of 65~mas. For example, a center offset of just 1/4 of a pixel corresponds to an apparent wavefront tilt of 16~mas, which over a 5.3~m baseline implies an excess photon path-length of 0.4~$\mu$m, or a phase of 30$^\circ$ at 4.8~$\mu$m, comparable to the phase gains in Figure~\ref{fig:Gains}. 

\begin{figure}[!htb]
\centering 
\centerline{\includegraphics[scale=0.25]{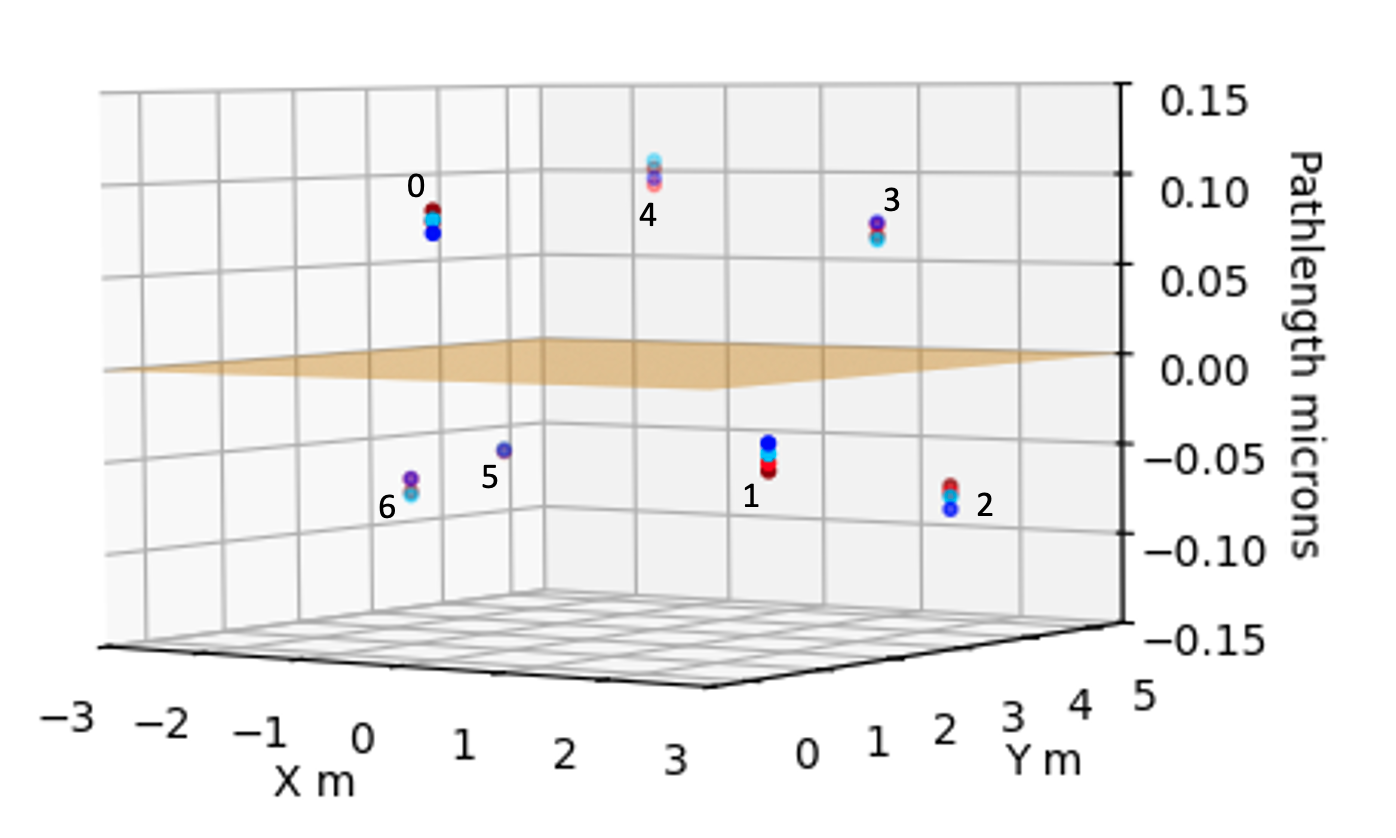}}
\caption{A 3-D projection of the residual static metrology measured for the JWST mirror segments using the AMI data. Each set of points corresponds to four measurements (3.8$\mu$m and 4.8$\mu$m for WR137 and HD 228337) of residual photon pathlength after subtracting the lowest order (tip-tilt) plane, for each aperture in the mask (Table~\ref{table:metro}). Apertures are numbered (see Figure~\ref{fig:piston}). The orange plane is the z=0 surface. 
}
\label{fig:metrology}
\end{figure}

After subtracting the best fit plane, Figure~\ref{fig:metrology} shows a 3D projection of derived path-length to each aperture for the four measurements (two sources and two wavelengths). These path-lengths correspond to the JWST mirror segment pistons. These values are listed in Table~\ref{table:metro}. Values range from about -80~nm to 100~nm, with the full scatter among the four independent measurements at each hole $\sim 10$~nm to 15~nm. 

\begin{figure}[!htb]
\centering 
\centerline{\includegraphics[scale=0.3]{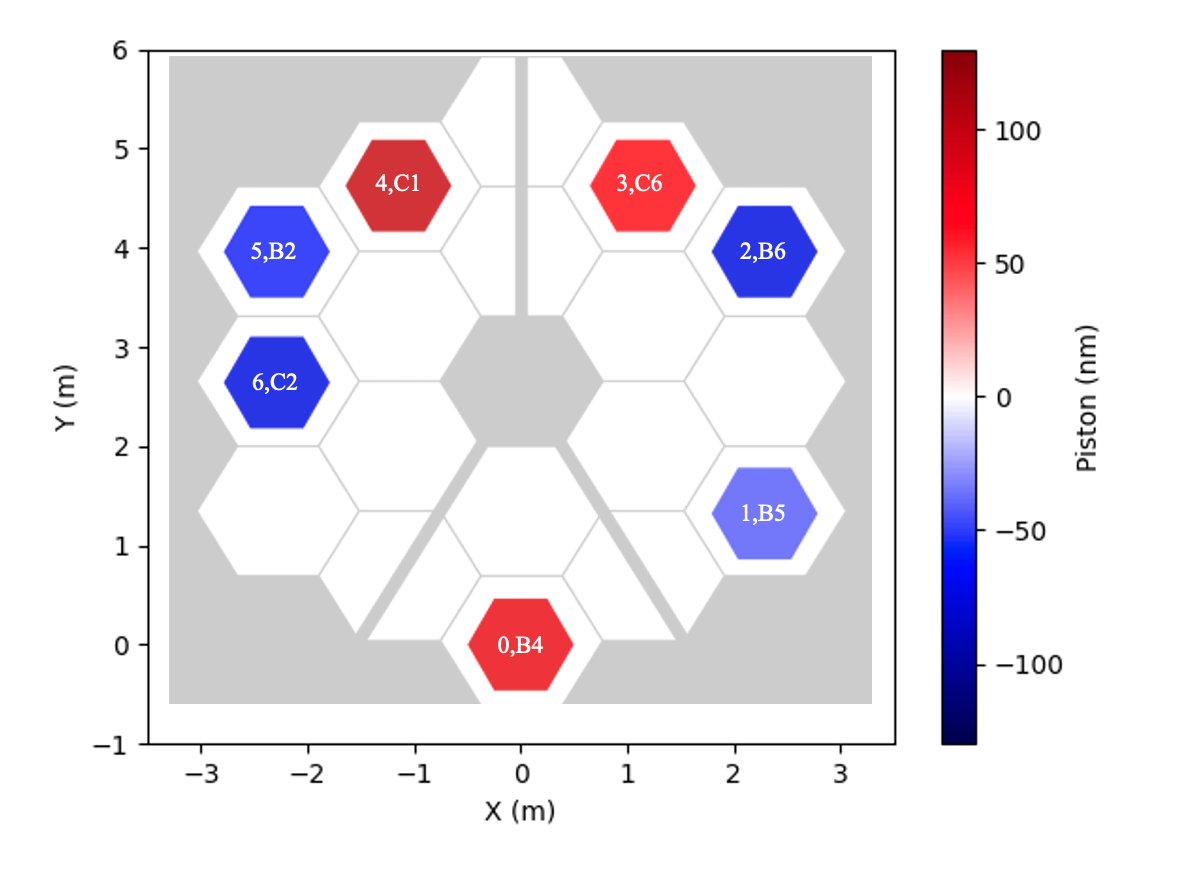}}
\caption{Face-on view of the measured piston values for the 7 AMI apertures from the phase elf-calibration process. Values are the mean of the four measurements shown for each aperture in Figure~\ref{fig:metrology} and listed in Table~\ref{table:metro}. Aperture designations are listed, including aperture numbers used herein and original JWST segment designations for reference. 
}
\label{fig:piston}
\end{figure}

Figure~\ref{fig:piston} shows the mean of these four measurements for each aperture in a face-on view, projected on the JWST mirror segment diagram. The pattern of positive and negative piston values for the segments, and the magnitude of the pistons, are comparable to the analysis in Desdoigt et al.\cite{desdoigts2025} (their figure 10). However, their more involved analysis, including an optics model for the system, allowed for derivation of phase gradients across each hole, and hence a direct comparison is difficult. Phase gradients across a given hole affect the primary beam response of that element. 

The measurement of segment pistons using self-calibration will be limited by non-closing phase errors. Our analysis of the point source calibrator HD 228337 shows there are non-closing phase errors which will corrupt baseline-based self-calibration solutions (Figure~\ref{fig:Closure}). The mean absolute value of the closure phases for all triads is $2.0^\circ$ with an rms scatter of $2^\circ$. Hence, the non-closing phase corruptions indicate an uncertainty level for the self-calibration derived pistons $\sim 27$~nm, using equ.~\ref{eqn:path}. This value is similar to the measured slow variations in the gains at the level of $\sim 2^\circ$, and to the full range in the four independent measurements of piston at each aperture $\sim$ 10~nm to 15~nm (Figure~\ref{fig:metrology}, Table~\ref{table:metro}). 

\begin{table}
\centering
\footnotesize
\caption{Piston Values}
\begin{tabular}{lcccccccc} 
  \hline
  \hline
  Object & Wavelength & 0  & 1 & 2 & 3 & 4 & 5 & 6 \\
~ & $\mu$m & nm & nm & nm & nm & nm & nm & nm  \\
\hline
HD228 & 4.8 & 82.5 & -41.0 &  -74.7 & 63.6  & 107.3 &  -58.2 &  -79.4 \\
HD228 & 3.8 & 87.4 & -50.0 &  -68.9 & 65.9 & 102.9 &  -60.0 &  -77.4 \\
WR137 & 4.8 & 75.5 & -35.4 & -81.7 &  73.3 & 97.2 & -59.1 & -69.9 \\
WR137 & 3.8 & 81.6 & -45.7 &  -72.5 & 72.1 & 93.4 & -58.7 & -70.2 \\
\hline
\hline
\vspace{0.1cm}
\end{tabular}
\label{table:metro}
\end{table}

\subsection{Baseline-based errors: phase}
\label{sec:visphs}

Our final analysis involves correcting for non-closing phases using the visibilities measured on the calibrator star HD 228337 (the visibility amplitudes were already normalized according to coherences measured on HD 228337; Figure~\ref{fig:Closure}). In this case, the visibility phases are corrected by subtracting the measured phases on HD 228337 from those measured for WR137. We then run through self-calibration to remove any residual gain phase or amplitude difference between the source and the calibrator.

The results for the image parameters are listed in Table~\ref{table:image}. The image rms improves at both 4.8~$\mu$m and 3.8~$\mu$m by about 23\%, and the peak off-source negative feature decreases by 40\% and 20\%, respectively.  Hence, the dynamic range has improved to $\sim 240$.  

\section{Discussion}
\label{sec:discussion}

We have demonstrated a new method of imaging with aperture masking interferometry data from NIRISS at the JWST. Using the visibilities derived from the Fourier transform of the interferograms, we employ the standard `self-calibration' process, in which an iterative joint optimization is performed for both the target source structure and the complex voltage gain responses of the interferometric elements.  This is a methodology long used in radio astronomy, but not, to date, in optical interferometry. By using target source data, the method represents essentially a 'real-time' wavefront error sensor.

The resulting images are consistent with results for these data derived using the standard AMI NIRISS processing algorithms\cite{lau2024}. These standard algorithms involve model fitting to closure phases and visibility amplitudes. Consistency between methods includes detection of the faint 300~mas extension of the source, at a similar contrast and extent in both reductions, and a similar image dynamic range.  

We consider the magnitude of non-closing, or baseline-based, amplitude and phase errors as well, through analysis of the closure phases and closure amplitudes on the calibrator star (Figure~\ref{fig:Closure}). Such errors will not be fixed by element based gain self-calibration. The visibility amplitudes on the calibrator star decrease substantially with baseline length, which is not expected for a point source, where visibility amplitudes should all be unity (Figure~\ref{fig:Closure}). Further, the closure amplitudes on the calibrator star differ from the expected value of unity by up to almost a factor 2 (Figure~\ref{fig:Closure}). The origin of these effects are likely some combination of under-sampling of the PSF, charge migration, and possibly near-field optics issues \cite{argyriou2023,desdoigts2024,desdoigts2025}. The magnitude of these corruptions requires that, before self-calibration, we make a baseline-based visibility amplitude correction using the coherences measured on the calibrator. 

The self-calibration phase gains correspond to photon path-lengths through the optical system to each hole, corresponding to the piston phases for the mirror segments. Piston values derived range from $-80$~nm to 100~nm, with a total scatter between the four measurements (two source and two wavelengths) of 10~nm to 15~nm for each hole. The pattern of positive and negative pistons for the mask holes, and the magnitudes, are comparable to those derived by Desdoigts et al.\cite{desdoigts2025}.

Concerning non-closing (baseline-based) phase errors, the pupil optics of the AMI system on JWST has potential near-field effects that could lead to closure errors. The measured closure phases on the calibrator star are small, but finite, with a mean departure from zero closure $\sim 2^\circ$, and a maximum of 6.5$^\circ$. While small, these errors are still substantial when considering high dynamic ranging imaging. Indeed, non-closing visibility phase errors at this level should limit the image dynamic range to about the observed level of $\sim 200$. Further, non-closing phases imply an uncertainty in the segment piston measurements derived from self-calibration at the level of $\sim 27$~nm.

As a final test, we made both baseline-based phase and amplitude corrections to the visibilities on WR137, using the visibility measurements on HD 228337. The resulting image dynamic range improves by about 23\%. 

Ultimately, high dynamic range imaging of sources with complex structure will be limited at the JWST-AMI system by the sparse u,v sampling, although the sampling may be improved by considering sub-structure across a u,v sample, or through rotation\cite{ford2014,Sivaramakrishnan2023,charles2025}. Still, for pushing to, or below, the diffraction limit at very high contrast for simple sources, the JWST-AMI system remains an important avenue to explore. 

\subsection{Acknowledgments}
We thank B. Pope and L. Desdoigts for interesting exchanges concerning AMI at the JWST, and Dr. Lau for permission to reproduce a figure. The National Radio Astronomy Observatory is a facility of the National Science Foundation operated under cooperative agreement by Associated Universities, Inc.. Image processing was performed using the Software: Astronomical Image Processing System (AIPS) \cite{Greisen2003} and Common Astronomical Software Applications (CASA) \cite{casa:2017}. 
Related patents and patent applications: Patent No. US 12,104,901 B2; Provisional Patent No. 63/355,174 (RL 8127.032.USPR); Provisional Patent No. 63/648,303 (RL 8127.306.USPR). Some/all of the data presented in this paper were obtained from the Multimission Archive at the Space Telescope Science Institute (MAST). STScI is operated by the Association of Universities for Research in Astronomy, Inc., under NASA contract NAS5-26555. Support for MAST for non-HST data is provided by the NASA Office of Space Science via grant NAG5-7584 and by other grants and contracts

\subsection*{Disclosures}
The authors declare that there are no financial interests, commercial affiliations, or other potential conflicts of interest that could have influenced the objectivity of this research or the writing of this paper.

{\bf Disclosure Statement:} The authors declare no conflicts of interest.

{\bf Data availability:} The JWST interferograms underlying the results presented in this paper are available on the MAST-JWST public archive.


\bibliography{report}   
\bibliographystyle{spiejour}   



\end{document}